\documentclass[aps,prl,twocolumn,showpacs,superscriptaddress,groupedaddress]{revtex4}  
\usepackage{graphicx}  
\usepackage{dcolumn}   
\usepackage{bm}        
\usepackage{amssymb}   
\usepackage{amsmath}
\usepackage{xcolor}
\usepackage{booktabs, multirow, array, makecell, caption}

\begin{document}

\widetext

\title{Description of the low-lying collective states of $^{96}$Zr based on the quadrupole-collective Bohr Hamiltonian}

\author{E.V. Mardyban}%
\author{E.A. Kolganova}
\address{Joint Institute for Nuclear Research, 141980 Dubna, Moscow region, Russia}
\address{Dubna State University, 141982 Dubna, Moscow Region, Russia}
\author{T.M. Shneidman}
\address{Joint Institute for Nuclear Research, 141980 Dubna, Moscow region, Russia}
\address{Kazan Federal University, Kazan 420008, Russia}
\author{R.V. Jolos}
\address{Joint Institute for Nuclear Research, 141980 Dubna, Moscow region, Russia}
\address{Dubna State University, 141982 Dubna, Moscow Region, Russia}
\author{N. Pietralla}
\address{Institut f\"ur Kernphysik, TU Darmstadt, Schlossgartenstr. 9, D-64289, Darmstadt, Germany}
\date{\today}

\begin{abstract}
\begin{description}
\item[Background:] Experimental data on $^{96}$Zr indicate  coexisting spherical and deformed structures with small mixing amplitudes. Several collective low-lying states and E2 and M1 transitions are observed for this nucleus. A consideration of these data in the full framework of the Geometrical Collective Model is necessary for $^{96}$Zr.
\item[Purpose:] To investigate the observed properties of the low-lying collective states of $^{96}$Zr  based on the Geometrical Collective Model.
\item[Method:] The quadrupole-collective Bohr Hamiltonian depending on both $\beta$ and $\gamma$ shape variables with a potential having spherical and deformed minima,  is applied. The  relative depth of two minima,  height and width of the barrier, rigidity of the potential  near both minima are determined so as to achieve a satisfactory description of the observed properties of the low-lying collective quadrupole states of $^{96}$Zr.
\item[Results:]  Good agreement with the experimental data on the excitation energies, $B(E2)$ and $B(M1; 2^+_2\rightarrow 2^+_1)$ reduced transition probabilities is obtained.
\item[Conclusion:] It is shown that the low-energy structure of $^{96}$Zr can be described in a satisfactory way within the Geometrical Collective Model with a potential function supporting shape coexistence without other restrictions of its shape. However, the  excitation energy of the $2^+_2$ state can be reproduced only if the rotation inertia coefficient is taken by four times smaller than the vibrational one in the region of the deformed well. It is shown also that shell effects are important for the description of the $B(M1; 2^+_2 \rightarrow 2^+_1)$ and $B(M1; 3^+_1 \rightarrow 2^+_1)$ transition probabilities. An indication for the influence of the pairing vibrational mode on the $0^+_2 \rightarrow 0^+_1$ transition is confirmed in agreement with the previous result.
\end{description}
\end{abstract}

\maketitle

\section{Introduction}

It is well known for a long time that nuclei can exhibit both spherical and deformed shapes including the intermediate region of nuclei transitional from spherical to deformed. What is more interesting is the phenomenon that a given nucleus can exhibit different shapes depending on the excitation energy. This phenomenon of shape coexistence has in recent years become the subject of many investigations in nuclear physics. Even more,  shape coexistence is becoming to be considered a near-universal property of nuclei \cite{Heyde1}. A large number of papers, including reviews \cite{Heyde1,Heyde2,Heyde3,Poves}, are devoted to investigation of shape coexistence \cite{Garcia,Togashi,Gavrilov,Sieja,Garcia2,Boyukata,Liu,Petrovici1,Rodriguez,Skalski1,Xiang,Mei,Skalski2,Ozen,Fortune,Buscher}. Various approaches have been employed to study this phenomenon \cite{Bender,Niksic,Federman,Heyde4,Etch,Holt,Niksic1}.

Among different examples of shape coexistence evidence for Zr isotopes with their change of the shape with excitation energy are especially interesting. Shape evolution can be characterized by a smooth or abrupt transition from spherical to deformed shape, and a significant or suppressed mixing of configurations with different shapes can take place. Such information is contained in electromagnetic transition probabilities and a high purity of coexisting shapes has been established in $^{96}$Zr \cite{Kremer}.

In this paper we apply the  Geometrical Collective-quadrupole Model to a description of the properties of the low-lying states of $^{96}$Zr including the shape coexistence phenomenon. Although an explanation of shape coexistence is a subject of microscopic nuclear modeling the Geometrical Collective Model deals directly with shape dynamical variables and, thus, may be capable of describing the dynamical consequences of shape coexistence and the properties of the collective low-lying states in general.

It is an open question whether a potential function in terms of shape variables can exist which allows for a reproduction of the data on the coexisting quadrupole collective structures of $^{96}$Zr.
The aim of the present paper is to investigate a possibility to describe, in principle,  the properties of the low-lying collective states of
$^{96}$Zr and the amount of mixing of the configurations characterized by spherical and deformed shapes based on the quadrupole collective Bohr Hamiltonian. It is also interesting in what characteristics of the collective states shell effects are most pronounced.

\section{Hamiltonian}

The quadrupole-collective Bohr Hamiltonian can be written as \cite{Niksic2}

\begin{widetext}
\begin{eqnarray}
\label{eq1}
H&=& -\frac{\hbar^2}{2 \sqrt{\omega r}} \left (\frac{1}{\beta^4} \left [ \frac{\partial}{\partial \beta}\sqrt{\frac{r}{\omega}}\beta^4 B_{\gamma \gamma} \frac{\partial}{\partial \beta}
-\frac{\partial}{\partial \beta}\sqrt{\frac{r}{\omega}}\beta^3 B_{\beta \gamma}\frac{\partial}{\partial \gamma}\right ]\right. \nonumber \\
&+&\left. \frac{1}{\beta \sin{3\gamma}} \left [ -\frac{\partial}{\partial \gamma} \sqrt{\frac{r}{\omega}}\sin{3\gamma}B_{\beta \gamma}\frac{\partial}{\partial \beta}+
\frac{1}{\beta}\frac{\partial}{\partial \gamma}\sqrt{\frac{r}{\omega}}\sin{3\gamma}B_{\beta \beta}\frac{\partial}{\partial \gamma}\right ] \right ) \nonumber \\
&+&\frac{1}{2} \sum_{k=1}^3 \frac{\hat J^2_k}{\Im_k(\beta)}+V(\beta, \gamma),
\end{eqnarray}
\end{widetext}
where $\omega = B_{\beta \beta}B_{\gamma \gamma}-B^2_{\beta \gamma}$ is the determinant of the vibrational inertia tensor
\begin{eqnarray}
\label{Bvib}
B_{vib} =
\begin{pmatrix}
B_{\beta \beta} & \beta B_{\beta \gamma} \\
\beta B_{\beta \gamma}& \beta^2 B_{\gamma \gamma}
\end{pmatrix}.
\end{eqnarray}
The moments of inertia $\Im_k$ with respect to the body-fixed axes are expressed as
\begin{eqnarray}
\label{moment_of_inertia}
\Im_k=4 B_k(\beta)\beta^2 \sin^2{\left (\gamma-\frac{2\pi k}{3}\right )}
\end{eqnarray}
and $r=B_1 B_2 B_3$. The components of the angular momentum in the body-fixed frame are denoted as $\hat J_k$ and can be expressed in terms of the  Euler angles.
The potential energy is denoted as $V(\beta,\gamma)$. The Hamiltonian of Eq.~(\ref{eq1}) is a general case of the conventional Bohr Hamiltonian \cite{Bohr1952} allowing for non-diagonal vibrational inertia.

In the present work we aim to investigate whether it is possible to construct a potential energy in such a way that all existing data on the energies of the lowest angular momentum excited
states and the transitions between these states will be described. If such a potential can be constructed will it demonstrate the shape coexistence by having two  minima: spherical and deformed. Previously in Ref.~\cite{Sazonov}, this problem was solved under the assumption that the $\gamma$ degrees of freedom can be separated from $\beta$ in the potential and the value of $\gamma$
is stabilized around $\gamma=0^\circ$. This is obviously a rather crude approximation at least in the region of small values of $\beta$.
In the present paper we avoid this assumption.

To simplify consideration, we make the following  assumptions for the inertia coefficients:
\begin{eqnarray}
\label{assumptions}
&& B_{\beta \beta}=B_{\gamma \gamma}=B_0, \hspace{10pt}  B_{\beta \gamma}=0,\nonumber \\
&& B_1(\beta)=B_2(\beta)=B_3(\beta)=b_{rot}(\beta) B_0,
\end{eqnarray}
where $B_0$ is the parameter scaling vibrational and rotational masses.
We keep in Eq.~(\ref{assumptions}) the rotational inertia coefficient because in the case of the well-deformed axially symmetric nuclei the inertia coefficient for the rotational motion is 4-10 times smaller than the inertia coefficient for the vibrational motion~\cite{Jolos1,Jolos2}. In a complete correspondence with this result it is shown below that in order to explain the excitation energy of the 2$^+_2$  state it is necessary to take $b_{rot}$ several times less than unity.

Under the assumptions of Eq.~(\ref{assumptions}), the Hamiltonian (\ref{eq1})  takes the form:
\begin{eqnarray}
\label{Hamiltonian_simplified_1}
\hat H &=& -\frac{\hbar^2}{2 B_0} \left (\frac{1}{b_{rot}^{3/2}}\frac{1}{\beta^4} \frac{\partial}{\partial \beta}\beta^4 b_{rot}^{3/2} \frac{\partial}{\partial \beta}
+\frac{1}{\beta^2 \sin{3\gamma}} \frac{\partial}{\partial \gamma}\sin{3\gamma}\right ) \nonumber \\
&+& \frac{1}{2} \sum_{k=1}^3 \frac{\hat J^2_k}{\Im_k(\beta)}+V(\beta, \gamma).
\end{eqnarray}

The potential energy $V(\beta, \gamma)$ is assumed to have  two minima, spherical and deformed, separated by a barrier. This is in correspondence with the considerations~\cite{Garcia,Gavrilov} 
in the interacting boson model with configuration mixing (IBM-CM) where two configurations with different total number of bosons have been taken into account in order to include the effect of shape coexistence. We expect that the wave function of the lowest excited states are localized  in these minima while the weight of the function inside the barrier region is strongly suppressed. Therefore, it is reasonable to assume that the quantity $b_{rot}$ has constant (but different) values in the regions of the spherical and deformed minima and the change from one value to another takes place in the region of the barrier. In this case, $b_{rot}$ can be taken outside of the derivative in Eq.~(\ref{Hamiltonian_simplified_1}) as it only gives the non-zero contribution in the barrier region where the wave function is close to zero. Thus, we obtain finally the following model Hamiltonian:
\begin{eqnarray}
\label{Hamiltonian_simplified_2}
H&=& -\frac{\hbar^2}{2 B_0} \left (\frac{1}{\beta^4} \frac{\partial}{\partial \beta}\beta^4 \frac{\partial}{\partial \beta}
+\frac{1}{\beta^2 \sin{3\gamma}} \frac{\partial}{\partial \gamma}\sin{3\gamma}\frac{\partial}{\partial \gamma}\right. \nonumber \\
&+& \left. \sum_{k=1}^3 \frac{\hat J^2_k}{4 b_{rot} \beta^2 \sin{\left (\gamma - \frac{2 \pi k}{3}\right )}}\right )+V(\beta, \gamma),
\end{eqnarray}
where
\begin{eqnarray}
\label{brot}
b_{rot}=
        \begin{cases}
            1 & \text{if $\beta \le \beta_m$, } \\
            b_{def}<1 & \text{if $\beta > \beta_m$.}
        \end{cases}
\end{eqnarray}
The magnitude of the $b_{rot}$ inside the deformed minimum is obtained by fitting the excitation energy of the $2^+_2$ state. The change from the spherical to deformed value of $b_{rot}$ occurs at $\beta =\beta_m$ which is taken around the maximum of the barrier separating spherical and deformed potential wells. Our calculations show that the precise value of $\beta_m$ does not affect the qualitative results of the calculations.

To solve the eigenvalue problem with the Hamiltonian (\ref{Hamiltonian_simplified_2})  we expand the eigenfunctions in terms of a complete set of basis functions
that depend on the deformation variables $\beta$ and $\gamma$ and the Euler angles. For each value of angular momentum $I$, the basis functions are written as
\begin{eqnarray}
\label{basis1}
\Psi_{IM}^{n_\beta v \alpha}=R^{(n_\beta, v)}(\beta)\Upsilon_{v \alpha I M}(\gamma, \Omega),
\end{eqnarray}
where $\Upsilon_{v \alpha I M}$ is the SO(5)$ \supset$ SO(3) spherical harmonics, which are the eigenfunctions of the operator $\hat \Lambda^2$:
\begin{widetext}
\begin{eqnarray}
\hat \Lambda^2 \Upsilon_{v \alpha I M} = \left [ -\frac{1}{\sin{3\gamma}} \frac{\partial}{\partial \gamma} \sin{3\gamma} \frac{\partial}{\partial \gamma} +
 \frac{1}{4} \sum_{k} \frac{\hat J_k^2}{\sin^2{(\gamma -\frac{2\pi k}{3})}} \right ]\Upsilon_{v \alpha I M} = v (v+3)\Upsilon_{v \alpha I M}.
\label{laplacian}
\end{eqnarray}
\end{widetext}
In addition to the angular momentum $I$ and its projection $M$, each function $\Upsilon_{v \alpha I M}$ is labeled by the SO(5) seniority quantum number $v$ and a multiplicity index $\alpha$,
 which is required for $v \ge 6$. In the following, both indices $v$ and $\alpha$ will be replaced by the running index $n_\gamma = 0, 1, 2, \dots$.

The $\Upsilon_{v \alpha I M}$ can be explicitly constructed as a sum over the states with explicit value of the projection $K$ of the angular momentum on the intrinsic axis
\cite{Rowe2004,Caprio2009}
\begin{eqnarray}
\Upsilon_{v \alpha I M}(\gamma, \Omega)=\sum_{K=0, even}^{I}F_{v \alpha I,K}(\gamma)\xi_{KM}^{I}(\Omega),
\end{eqnarray}
where
\begin{eqnarray}
\label{eq11}
\xi_{KM}^{I}(\Omega)=\frac{1}{\sqrt{2(1+\delta_{K0})}}\left [D^I_{M\  K}(\Omega)\right.\\
\left.
+(-1)^I D^I_{M\  -K}(\Omega)\right ]
\end{eqnarray}
 and the $F_{v \alpha I,K}(\gamma)$ are polynomials constructed from the trigonometrical functions of $\gamma$ \cite{Rowe-Wood}.

The basis wave functions $R^{(n_\beta, v)}$  are chosen as the eigenfunctions of the harmonic oscillator Hamiltonian in $\beta$:
\begin{eqnarray}
h_{h.o.}=\frac{1}{2}\left(-\frac{1}{\beta^4}\frac{\partial}{\partial\beta}\beta^4\frac{\partial}{\partial\beta}+\frac{v(v+3)}{\beta^2}+\frac{\beta^2}{\beta_0^4}\right).
\label{oscillator}
\end{eqnarray}

The eigenfunctions of $h_{h.o.}$ have the following analytical form
\begin{eqnarray}
\label{sol_1}
R_{n_\beta, v}(\beta)=N_\beta \left (\frac{\beta}{\beta_0}\right) ^v L_{n_\beta}^{v+3/2}\left (\frac{\beta^2}{\beta_0^2} \right ) \exp{\left ( -\frac{\beta^2}{2\beta_0^2} \right )},
\label{basis_beta}
\end{eqnarray}
where $\beta_0$ is an oscillator length and the normalization constant  $N_\beta$ is given as:
\begin{equation}
\label{N_const}
N_\beta=\sqrt{ \frac{2n_\beta!}{\Gamma(n_\beta+v+5/2)} }.
\end{equation}

The basis  functions $R_{n_\beta, v}$ are completely specified by the choice of the  oscillator length $\beta_0$. Our calculations have shown that the
 fastest  convergence of the results is obtained when $\beta_0$ is chosen to be equal to the value at the region of the barrier separating spherical and deformed minima so that the oscillator potential coincides with the potential $V(\beta,\gamma=0)$ at the top of the barrier.  For such a choice of $\beta_0$, $(n_\beta)_{max} = 30$ is enough to provide a convergence.

Diagonalization of the Hamiltonian \eqref{Hamiltonian_simplified_2} is realized in the basis of SO(5)-SO(3) spherical harmonics  $\Upsilon_{v \alpha I M}(\gamma, \Omega)$ truncated to some
maximum  seniority $v_{max}.$ As shown in \cite{Caprio2011}, taking $v_{max}=50$ is sufficient to provide a convergence of the calculation.
A concrete realization of the construction of  $\Upsilon_{v \alpha I M}(\gamma, \Omega)$  performed in \cite{Rowe2004,Caprio2009} is used in the present work. These functions
were first constructed in analytic form in \cite{Bes1959} for $I \le 6$.

The potential energy $V(\beta,\gamma)$ in \eqref{Hamiltonian_simplified_2} is chosen in the form
\begin{eqnarray}
V(\beta,\gamma)=U(\beta) +C_\gamma \beta^3(1-\cos{3\gamma}).
\label{potential energy}
\end{eqnarray}
In  \eqref{potential energy}, the deformed minimum of the potential energy is  localized around $\gamma=0$ for positive $C_{\gamma}$ as  it was assumed in our previous paper \cite{Sazonov}. At the same time, this form of $\gamma$-dependence of $V(\beta,\gamma)$ provides very weak $\gamma$-dependence of $V(\beta,\gamma)$ at small $\beta$ because of the factor $\beta^3$. The form of the potential energy at $\gamma=0$ ($U(\beta)$) and the parameter $C_\gamma$ which determines the stiffness of the potential with respect to $\gamma$ in the deformed minimum are fitted to reproduce
the experimental data. As the first step, we have taken $U(\beta)$ as it was numerically determined in \cite{Sazonov}, $B_{0}$=0.004 MeV$^{-1}$ and $b_{rot}=0.2$ and performed calculations with different values of $C_{\gamma}$. We have found that $C_{\gamma}$=50 MeV produces a reasonable value of the frequency of $\gamma$-vibrations close to 1.5 MeV. No significant changes were found in the calculation results for the excitation energies and the E2 transition probabilities when $C_{\gamma}$ was varied around 50 MeV.

As before in \cite{Sazonov}, to describe the shape of the axially-symmetric part of the potential we defined several points fixing the positions of the spherical and deformed minima, the rigidity of the potential near its
minima, and the height and width of the barrier separating  two minima. The deformation at the second minimum has been taken to be $\beta =$ 0.24 in agreement with the experimental
value of $B(E2; 2^+_2 \rightarrow  0^+_2 )$. The potential energy as a function of $\beta$ is determined by using a spline interpolation between selected points.
Then we solve numerically  the Schr\"odinger equation with Hamiltonian \eqref{Hamiltonian_simplified_2}, varying positions of the selected  points in order to get a satisfactory description of the energies
of the $2^+_1$ and $2^+_2$ states and the following transition probabilities: $B(E2; 2^+_2 \rightarrow 0^+_2)$, $B(E2; 2^+_1 \rightarrow 0^+_1)$, and $B(E2; 2^+_2 \rightarrow 0^+_1)$.
The number of points is taken to be 16 to provide a smooth change of the potential. However, not all the points are of the same physical importance. In principle, the number of points can be
 minimized as, obviously, the only relative depths of the minima and the height and width of the barrier leads to physically meaningful changes.
 The mass parameter  has been taken finally as $B_0=0.005$ MeV$^{-1}$ to fix the energy of the $0^+_2$ state.
\begin{figure}[hbt]
\includegraphics[width=0.4\textwidth]{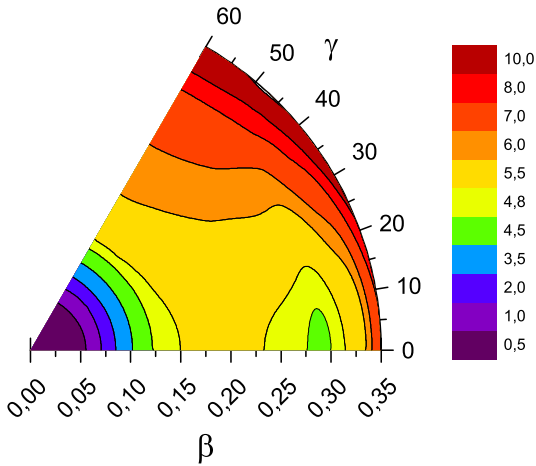}
\caption{\label{f1} Potential energy $V(\beta,\gamma)$ obtained in the calculations.}
\end{figure}

The resulting potentials $V(\beta,\gamma)$ is presented in Fig.~\ref{f1}. It is interesting that the inclusion of $\gamma$  as a dynamical variable
 leads to a significant change of the  shape of the  potential in comparison to the case when $\gamma$ was treated as a constant and not as a variable.
 The most important change occurs at the region of small $\beta$ where the potential becomes shallower. In this region, the resulting potential is practically independent on $\gamma$ and the
 wave function of the $0^+_1$ state becomes independent on $\gamma$ as well. This is not the case if $\gamma$ is treated as a constant. This lack of the phase space
 results in  the necessity to take a much deeper potential at small values of $\beta$ to hold the wave function of $0^+_1$ state inside the spherical minimum when $\gamma$ is not considered dynamic.

 \section{Results}

The Hamiltonian eigenfunctions $\Psi_{InM}$, where $I$ is the angular momentum, $M$ is its projection and $n$ is a multiplicity index, are obtained in calculations as a series expansions in the basic functions (\ref{basis1}). However, for discussions below it is more convenient to present them in the basis of functions $\xi^I_{KM}$ (\ref{eq11}):
\begin{eqnarray}
\label{xi}
\Psi_{InM}=\sum_K\psi_{InK}(\beta,\gamma)\frac{1}{\sqrt{2(1+\delta_{K0})}}
\left(D^I_{MK}(\Omega)\right.\nonumber\\
\left.+(-1)^ID^I_{M-K}(\Omega)\right)
\end{eqnarray}
We are using below the one-dimensional probability distributions over $\beta$ which are obtained by integration of 
$|\Psi_{InM}|^2$ over $\gamma$ and Euler angles
\begin{eqnarray}
\Phi_{In}(\beta) = \beta^4 \int_{0}^{\pi/3}\sin{3 \gamma}d\gamma \int d\Omega  |\Psi_{InM}|^2.
\label{probability}
\end{eqnarray}
and the weights of the wave functions in the spherical minimum $W_{In}$ determined as
\begin{eqnarray}
 W_{In} = \int_0^{\beta_m}  d\beta \Phi_{In}(\beta),
 \label{weights}
\end{eqnarray}
where $\beta_m$ is the position of the maximum of the barrier for $\gamma=0$.

\begin{figure}
a)
\includegraphics[width=0.2\textwidth]{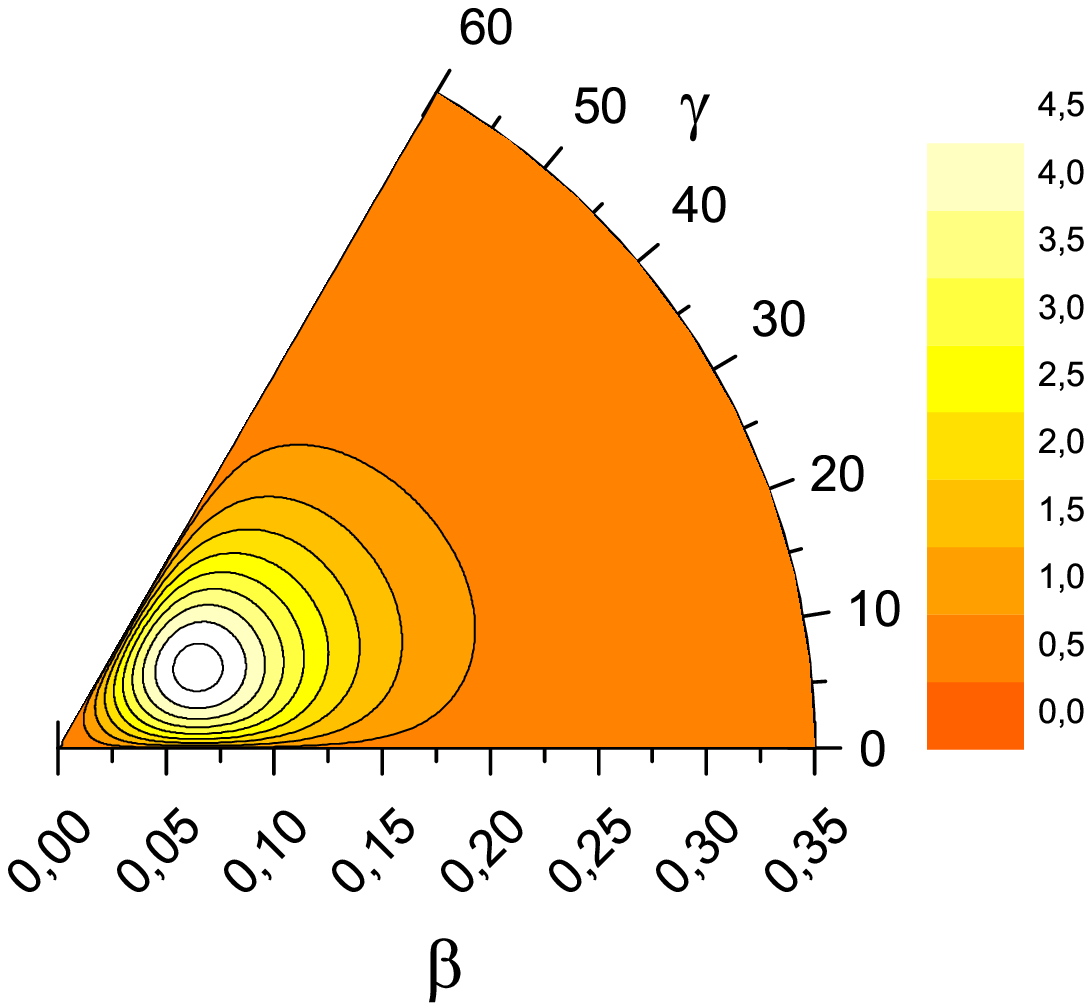}
b)
\includegraphics[width=0.2\textwidth]{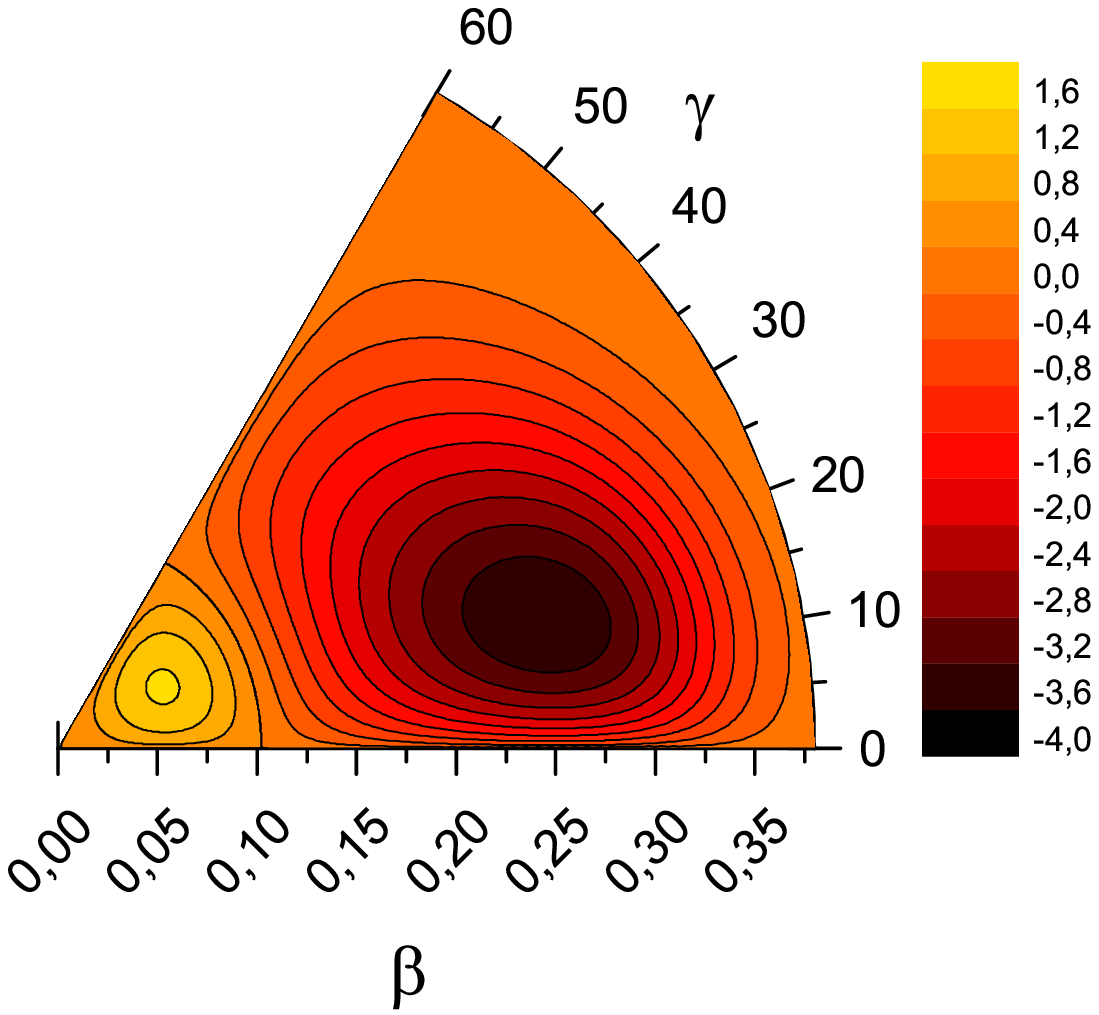}
\caption{\label{f34}Wave function of the $0^+_1$ (a) and $0^+_2$ (b) states.}
\end{figure}

\begin{figure}[hbt]
\centering
\includegraphics
[width=0.45\textwidth]{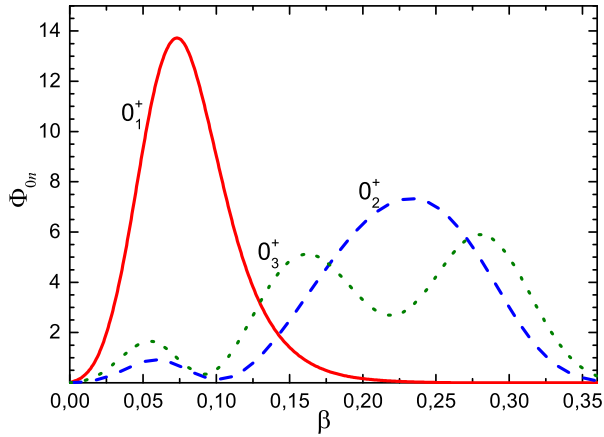}
\caption{\label{f5} Distribution over $\beta$
of the squares of the wave functions of the $0^+_1$ (solid line), $0^+_2$ (dashed line) and $0^+_3$ (dotted line) states calculated according (\ref{probability}).}
\end{figure}

The calculated wave function of the $0^+_1$ and $0^+_2$ states multiplied by the $\beta$ and $\gamma$- dependent volume element are presented in Fig.~\ref{f34}.
As it is seen, the wave function of the $0^+_1$ is strongly localized in the spherical minimum. The wave function of the $0^+_2$ state is mainly localized in the deformed minimum.
Their spherical weights are $W_{0_1}$=0.985 and  $W_{0_2}$=0.136 for  $0^+_1$ and $0^+_2$ states, respectively. The one-dimensional probability distribution over 
$\beta$ which can be obtained by integrating  $|\Psi_{IM}^{n_\beta v \alpha}|^2$  over $\gamma$ and the Euler angles are presented in Fig.~\ref{f5} for the $0^+_1$ and $0^+_2$ states.

For the lowest 2$^+$ states the situation is similar. The $2^+_1$ state  is localized in the spherical minimum with  the weight $W(2^+_1)$=0.928, while the second excited $2^+$ state is
only weakly presented there with $W(2^+_2)$ =0.144.

\begin{figure}[hbt]
a)
\includegraphics[width=0.2\textwidth]{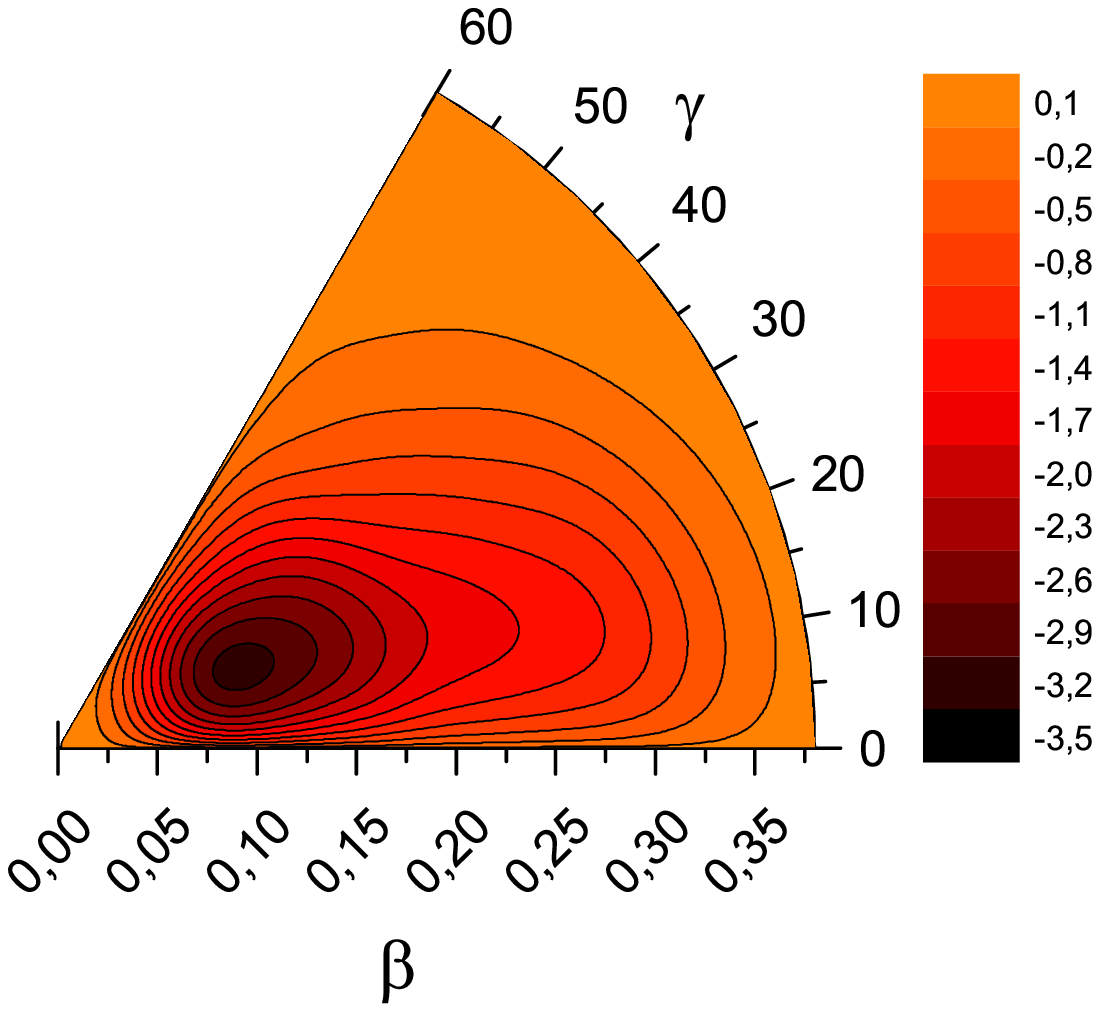}
b)
\includegraphics[width=0.2\textwidth]{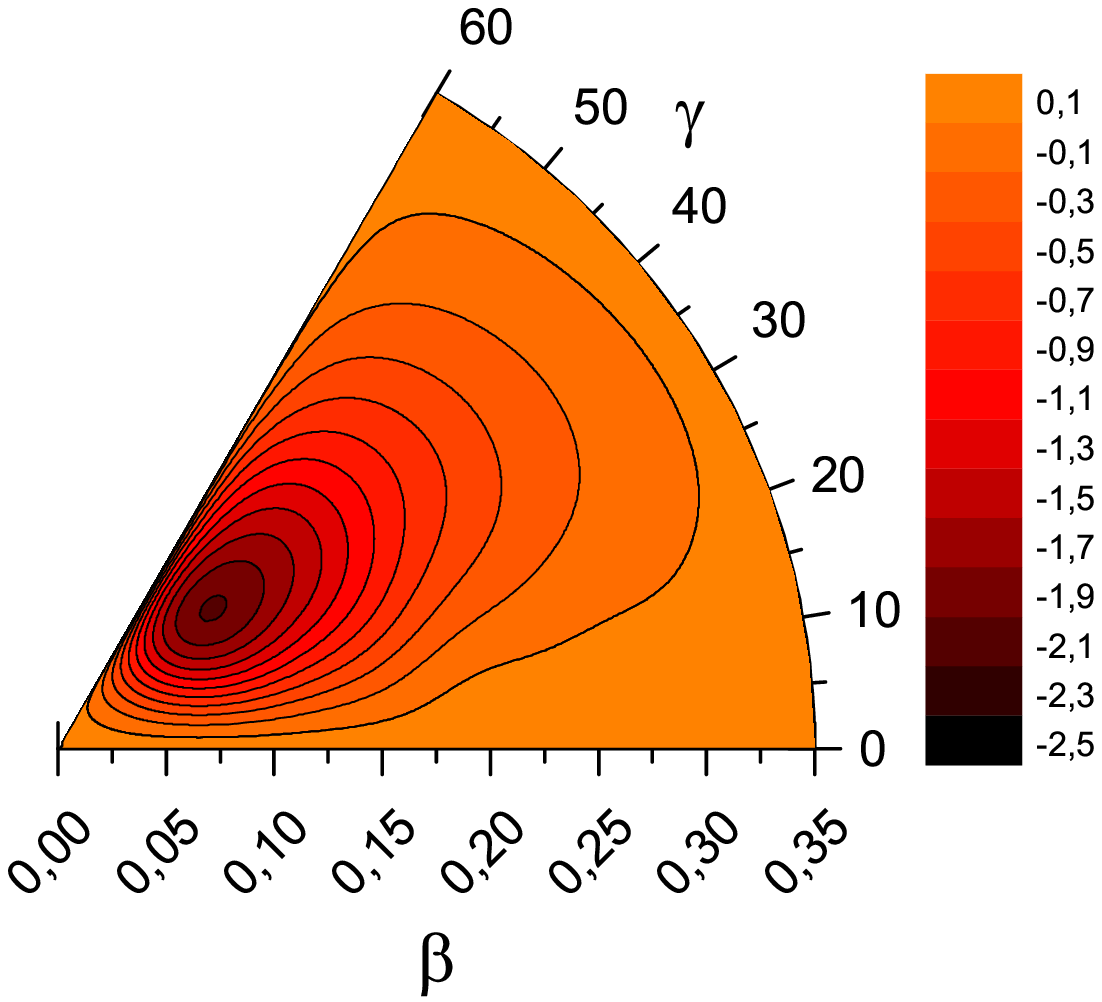}
\caption{\label{f67}The components of the wave function of the $2^+_1$ state with $K=0$ (a) and $K=2$ (b) multiplied by the volume element.}
\end{figure}

\begin{figure}[hbt]
a)
\includegraphics[width=0.2\textwidth]{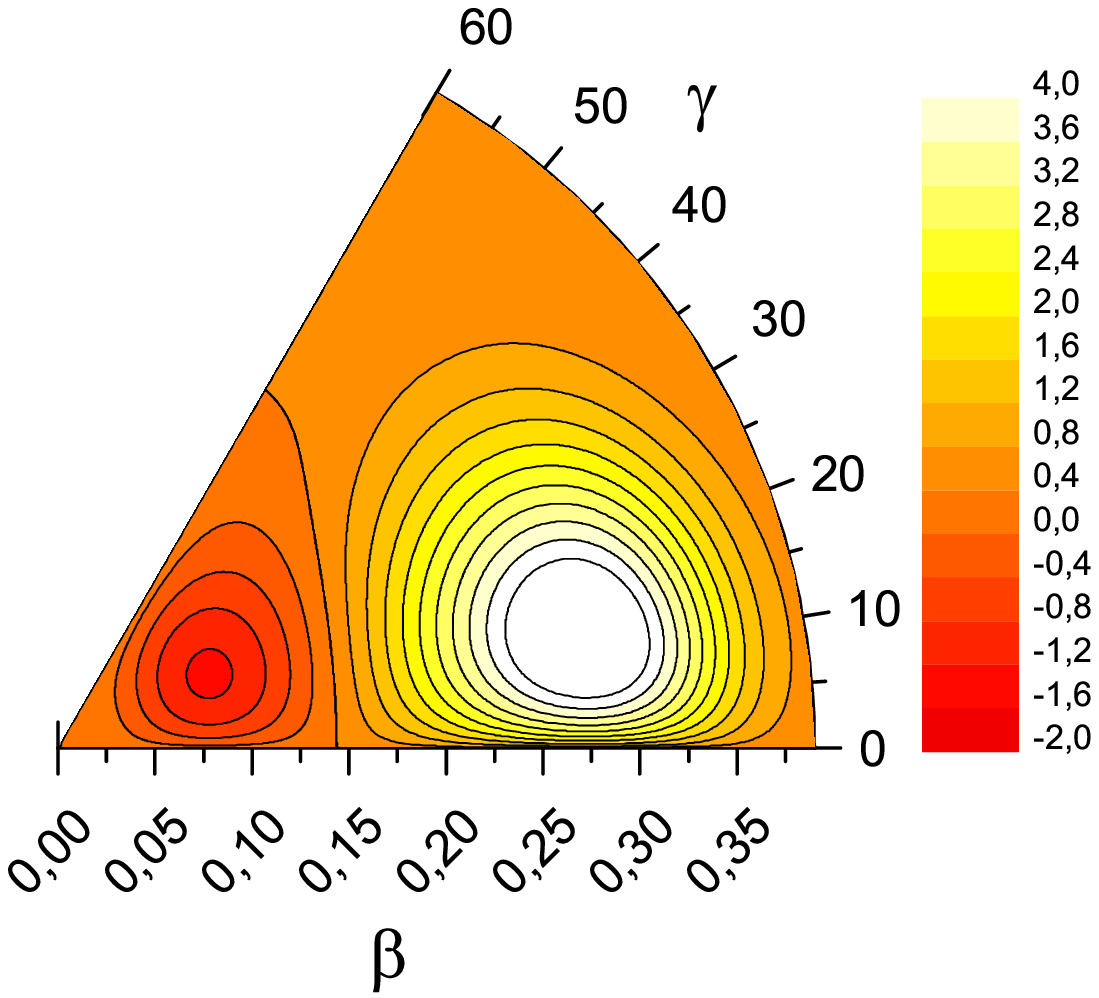}
b)
\includegraphics[width=0.2\textwidth]{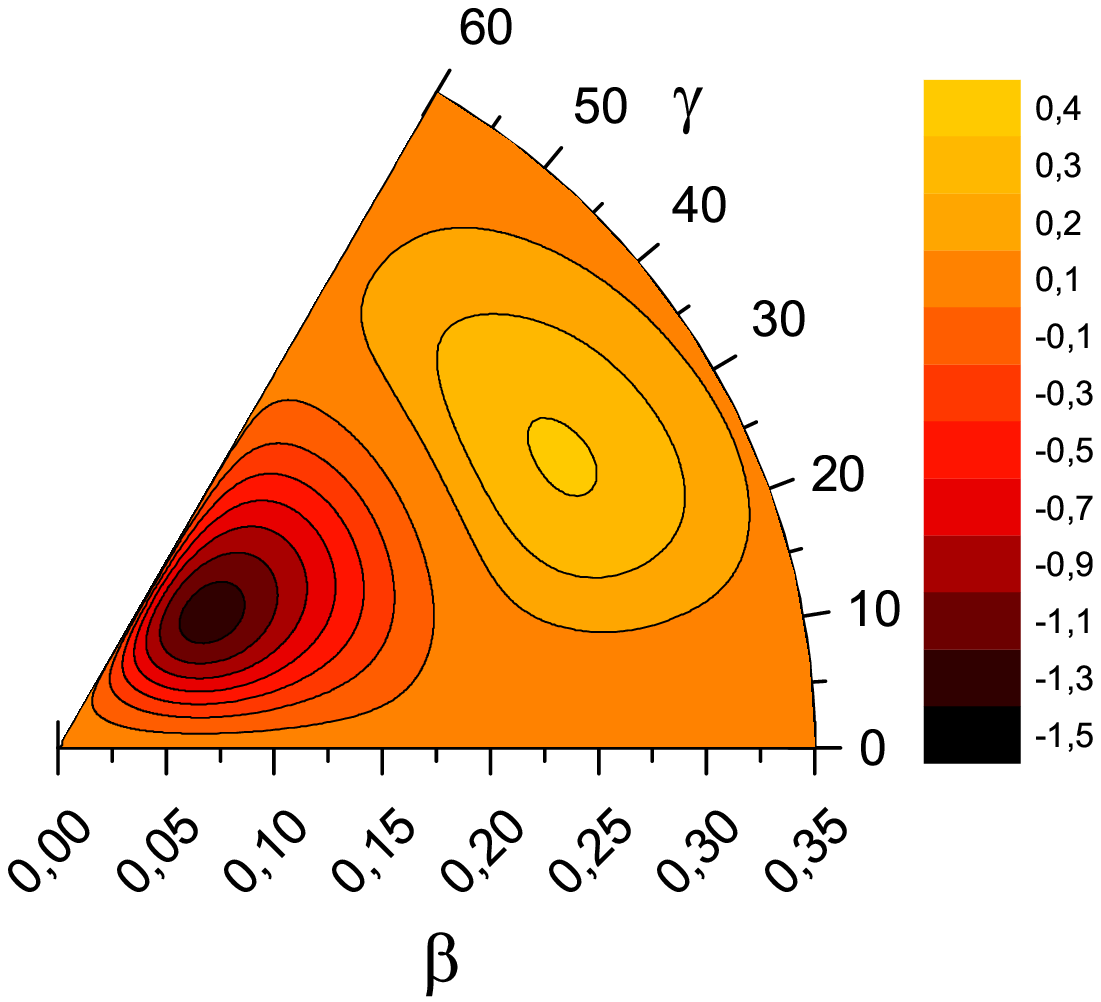}
\caption{\label{f89} The components of the wave function of the $2^+_2$ state with $K=0$ (a) and $K=2$ (b) multiplied by the volume element.}
\end{figure}

The wave functions of the $2^+$ states have components with $K$=0 and $K$=2  determined by the expansion
  (\ref{xi}).
The functions $\psi_{2^+nK}$ for $K$=0 and $K$=2 multiplied by the volume element are presented in Fig.~\ref{f67} for the $2^+_1$ state and in Fig.~\ref{f89} for the $2^+_2$ state.

Using these wave functions the matrix elements of an arbitrary operator $\hat F$ can be calculated as
\begin{eqnarray}
<f|\hat F| i> = \int_0^{\infty} \beta^4 d\beta \int_{0}^{\pi/3}\sin{3 \gamma}d\gamma \int d\Omega \Psi_{f}^* \hat F \Psi_{i}.
\label{matrix_element}
\end{eqnarray}
We are particularly interested in calculations of the $E2$ and $M1$ transition probabilities. The collective quadrupole operator responsible for $E2$ transitions is taken in the form
\begin{eqnarray}
Q^{coll}_{2 \mu}
&=& \frac{3Ze}{4\pi}R_0^{2} \left(  \beta \cos{\gamma} \phantom{\frac{1}{1}}  \! \! \!  D^{2}_{\mu 0}(\Omega) \right. \nonumber\\
 & + &  \left. \frac{1}{\sqrt{2}} \beta \sin{\gamma} \left( D^{2}_{\mu 2}(\Omega) + D^{2}_{\mu -2}(\Omega) \right) \right ) ,
\label{Q2operator}
\end{eqnarray}
where $R_0$ is the  equivalent volume-conserving spherical radius  of the nucleus and $Z$ is the nuclear charge number. The E0 transition strength $\rho^2(0^+_2\rightarrow 0^+_1)$ is calculated using the expression
\begin{eqnarray}
\label {rho}
\rho^2(0^+_2\rightarrow 0^+_1)=\left(\frac{3Ze}{4\pi}\right)^2|\langle0^+_2|\beta^2|0^+_1\rangle|^2
\end{eqnarray}
For the $M1$ transition operator we use the same  expression as in \cite{Sazonov}
\begin{eqnarray}
(M1)_\mu=\mu_N\sqrt{\frac{3}{4\pi}}g_R(\beta)I_\mu,
\end{eqnarray}
where $\mu_N$ is the nuclear magneton and $g_R(\beta)$ is the deformation-dependent collective $g$ factor.

The results of calculations for the energies of the low-lying states and the electromagnetic transition probabilities   are presented in Table~\ref{table:one} and Table~\ref{table:two} together with the available experimental data.
\begin{table}[tbh]
\centering
\setlength\aboverulesep{0pt}\setlength\belowrulesep{0pt}
\setcellgapes{3pt}\makegapedcells
\caption{The calculated and the experimental energies of the low-lying 0$^+$, 2$^+$, 3$^+$ and 4$^+$ states. The experimental data are taken
from \cite{Garcia,nndc1}}.
\label{table:one}
\begin{tabular}{l|c|c}
\hline
State  &    $E_{calc}$ (MeV)  & $E_{exp}$ (MeV)  \\
\hline
$E(0^+_2)$ &    1.582   & 1.582             \\
$E(0^+_3)$ &    2.443   &  2.695             \\
$E(0^+_4)$ &    3.049    &  2.926             \\
$E(2^+_1)$ &    1.724    & 1.750               \\
$E(2^+_2)$ &    2.236    & 2.226               \\
$E(2^+_3)$ &    2.974    & 2.669             \\
$E(2^+_4)$ &    3.338    & 3.249             \\
$E(3^+_1)$ &    2.653    & 2.439             \\
$E(4^+_1)$ &    2.983    &  2.857            \\
$E(4^+_2)$ &    3.447    &  3.082           \\

\hline
\end{tabular}
\end{table}

\begin{table}[tbh]
\centering
\setlength\aboverulesep{0pt}\setlength\belowrulesep{0pt}
\setcellgapes{3pt}\makegapedcells
\caption{The calculated and the experimental values of the electromagnetic transition probabilities in $^{96}$Zr. B(E2) values are given in W.u., B(M1) - in nuclear magnetons. The value of $Q(2^+_2)$ is given in $e\cdot barn$. Experimental data are taken from  \cite{nndc1,Witt}.}
\label{table:two}
\begin{tabular}{l|c|c}
\hline
transitions  &    calc  & exp   \\
\hline
$B(E2; 2^+_1 \rightarrow 0^+_1)$ &    5.23 & 2.3(3)      \\
$B(E2; 2^+_2 \rightarrow 0^+_1)$ &    0.39 & 0.26(8)   \\
$B(E2; 2^+_2 \rightarrow 0^+_2)$ &    26.0 & 36(11)   \\
$B(E2; 2^+_2 \rightarrow 2^+_1)$ &    6.49 & 2.8$_{-1.0}^{+1.5}$   \\
$B(E2; 3^+_1 \rightarrow 2^+_1)$ &    0.22 & 0.1$_{-0.1}^{+0.3}$   \\
$B(E2; 3^+_1 \rightarrow 2^+_2)$ &    4.26 &  -  \\
$B(E2; 0^+_3 \rightarrow 2^+_1)$ &    1.14 &  -    \\
$B(E2; 0^+_3 \rightarrow 2^+_2)$ &    69.8 & 34(9)    \\
$B(E2; 2^+_3 \rightarrow 2^+_1)$ &    10.6 & 50(70)    \\
$B(E2; 2^+_3 \rightarrow 2^+_2)$ &    1.85 &  $< 400$   \\
$B(E2; 4^+_1 \rightarrow 2^+_1)$ &    16.7 & 16$_{-13}^{+5}$   \\
$B(E2; 4^+_1 \rightarrow 2^+_2)$ &    43.0 & 56(44)    \\
$B(E2; 4^+_1 \rightarrow 3^+_1)$ &    7.59 &  -    \\
$B(E2; 0^+_4 \rightarrow 2^+_1)$ &    0.36 & 0.3(3)   \\
$B(E2; 0^+_4 \rightarrow 2^+_2)$ &    2.02 & 1.8(14)   \\
$B(E2; 4^+_2 \rightarrow 2^+_1)$ &    4.82 &  -  \\
$B(E2; 4^+_2 \rightarrow 2^+_2)$ &    16.6 &  -   \\
$B(E2; 4^+_2 \rightarrow 3^+_1)$ &    0.07 &  -  \\
$B(E2; 2^+_4 \rightarrow 2^+_1)$ &    2.53 &  -   \\
$\rho^2(0^+_2 \rightarrow 0^+_1)$&    0.0023 &  0.0075 \\
$\rho^2(0^+_3 \rightarrow 0^+_1)$&    0.001 & 0.004  \\
$\rho^2(0^+_3 \rightarrow 0^+_2)$&    0.038 & 0.0035  \\
$B(M1; 2^+_2 \rightarrow 2^+_1)$ &    0.071 &  0.14(5)    \\
$B(M1; 3^+_1 \rightarrow 2^+_1)$ &    0.0002 &  0.3(1)    \\
$Q(2^+_2)$                       &    $-0.5$ &   -     \\
\hline
\end{tabular}
\end{table}

As it is seen from the results presented in Tables \ref{table:one} and \ref{table:two} the agreement between the calculated results and the experimental data is quite satisfactory.
This applies not only to the $0^+_2$, $2^+_1$ and $2^+_2$ states on which attention was focused primarily in determining the form of the collective potential. 

\begin{figure}[hbt]
\includegraphics[width=0.45\textwidth]{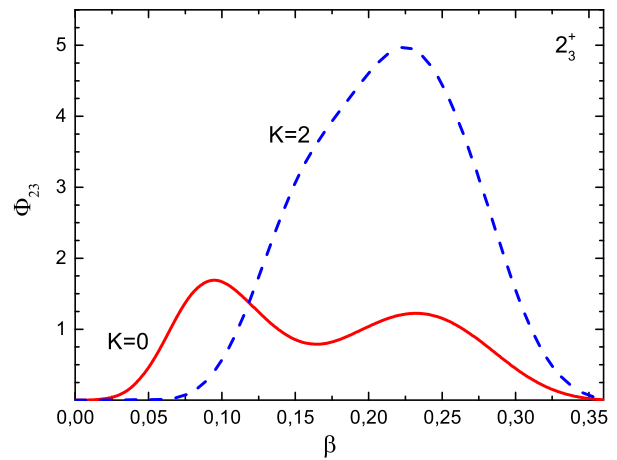}
\caption{\label{f10} Distribution over $\beta$ of the squares of the components of the $2^+_3$ state with $K$=0 (solid line) and $K$=2 (dashed line) calculated according (\ref{probability}).}
\end{figure}

Let us consider the results for the $0^+_3$, $0^+_4$, $2^+_3$ and $4^+_1$ excited states.
The calculated energy of the $2^+_3$ state exceeds the experimental value by 300 keV which is 10\% of the total excitation energy of this state. The calculated value of B(E2; $2^+_3\rightarrow 2^+_1$)=10.6 W.u. is quite collective as the experimental result. The experimental value~\cite{nndc1} can vary between 0 and 120 W.u. depending on the quite uncertain lifetime of this level and on the unknown multipolarity of its decay transition to the $2^+_1$state.
A distribution of the wave function of the $2^+_3$ state over $\beta$, determined by (\ref{probability}), is presented in Fig.~\ref{f10}. It is seen that the component with $K$=0 is almost equally distributed  between the spherical and deformed minima. The component with $K$=2 is predominantly located in the deformed minimum.

The experimental value of the excitation energy of the $3^+_1$ state and the value of B(E2; $3^+_1\rightarrow 2^+_1$) are reproduced by the calculations quite well. However, the experimental value of B(M1; $3^+_1\rightarrow 2^+_1$)=0.3 $\mu_N^2$ is too large to be reproduced in the framework of the collective model. For instance, the value of B(M1; $3^+_{\gamma}\rightarrow 2^+_{\gamma}$) for transition between the states of the $\gamma$-band in $^{168}$Er is equal to 0.003 $\mu_N^2$  only, i.e. two orders of magnitude less than the value for $^{96}$Zr. It could mean that the $3^+_1$ state of $^{96}$Zr has a large component of the shell model neutron configuration $(s^1_{1/2}d^{-1}_{5/2})_3$ or even its structure is almost exhausted by this configuration \cite{Witt}. We mention, however, that the experimental value of B(E2; $3^+_1\rightarrow 2^+_1$) can be reproduced only if both states have a collective admixture, since for the explanation of the experimental 
B(E2; $3^+_1\rightarrow 2^+_1$) value the shell model neutron configurations $(s^1_{1/2}d^{-1}_{5/2})_{2,3}$ requires a neutron E2 effective charge equal to one. The calculated wave function of the $3^+_1$ state is almost completely localized in the deformed minimum: $W_{3_1}$=0.96.

The strong E2 transition between the $0^+_3$ and the deformed $2^+_2$ states is reproduced by our calculations because a significant part of the wave function of the $0^+_3$ state is localized in the  deformed minimum of the potential (see Fig.\ref{f5}).

It is indicated in \cite{Witt} that the $4^+$ states at 2750 keV and 2781 keV   presented in \cite{nndc1}  have been observed in one experiment each only and were never been confirmed. For this reason we disregard these states and compare the calculated characteristics of the $4^+_1$ state with the experimental data for the $4^+$ state observed at 2857 keV.

\begin{figure}[hbt]
\centering
\includegraphics[width=0.45\textwidth]{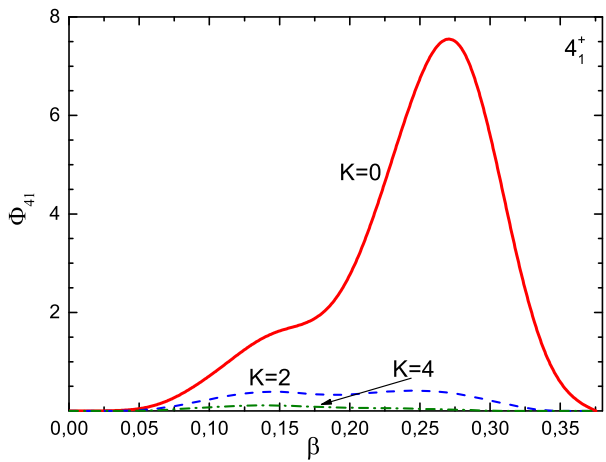}
\caption{\label{f12} Distribution over $\beta$ of the squares of the components of the $4^+_1$ state with $K$=0 (solid), 2 (dashed) and 4 (dot-dashed) calculated according (\ref{probability}).}
\end{figure}

Our calculations reproduce the value of the very collective E2 transition $4^+_1\rightarrow 2^+_2$ which shows that the significant part of the wave function of the $4^+_1$ state is localized in the deformed minimum. This fact is confirmed by the distribution of the wave function of the $4^+_1$ state shown in Fig.~\ref{f12}. It is seen also that the wave function of the $4^+_1$ state is exhausted by the $K$=0 component. The calculated ratio
B(E2; $4^+_1\rightarrow 2^+_2$)/B(E2; $2^+_2\rightarrow 0^+_2$) = 1.65
is close to the Alaga value 1.43 for axially deformed nuclei. A distribution of the $K$=0, 2 and 4 components of the wave function of the $4^+_1$ state  indicates that the large part of the total wave function is indeed located in the deformed minimum. At the same time the calculated B(E2; $4^+_1\rightarrow 2^+_1$) value agrees within the limit of the experimental error  with the observed value. The calculated  ratio $\left(E(4^+_1)-E(0^+_2)\right)$/$\left(E(2^+_2)-E(0^+_2)\right)$ is equal to 2.14 which is close to the spherical  limit. The experimental value of this
ratio 1.98 practically coincides with the value for the spherical harmonic oscillator.

This astonishing apparent correspondence of the $4^+_1$ state's
properties to contradicting limits of the collective model can be
understood from the following consideration.
The dominant parts of the wave functions of the $2^+_2$ and $4^+_1$ states are
located in the deformed minimum. However, smaller parts  of the wave
functions of these states are spread over the spherical minimum. This
fact allows us to consider the $2^+_2$ and $4^+_1$ states  as a mixture of
the two dominant, lowest-lying spherical and deformed components, each.
As a result of this mixing, the $4^+_1$ state with dominantly deformed
character is shifted down in energy because it is the lowest $4^+$ state.
At the same time, the predominantly deformed $2^+_2$ state is shifted up in
energy since it is the second excited $2^+$ state.
This lowering of the excitation energy of the 4+ state and this increase
of the $2^+$ state's energy in the deformed well leads to the observed significant
reduction of the $R_{4/2}$ ratio from the value of $10/3$ expected for
axially-deformed nuclei towards a smaller value closer to 2.

Let us analyze the result obtained for $\rho^2(0^+_2 \rightarrow 0^+_1)$ which is by factor 3 smaller than the experimental value.
The definition of the $\rho^2(0^+_2\rightarrow 0^+_1)$ value is given in (\ref{rho}).
In order to get an expression for $\langle 0^+_2 | \beta^2 | 0_1 \rangle$ in terms of the  quantities whose values are known from other experiments let us calculate the double commutator $[[H, \beta^2], \beta^2]$ using the Hamiltonian~(\ref{Hamiltonian_simplified_1}). The result is
\begin{eqnarray}
\label{eq:eleven}
\left[ \left[ H, \beta^2 \right], \beta^2 \right] = \frac{4\hbar^2}{B_0} \beta^2.
\end{eqnarray}
Taking the average of~(\ref{eq:eleven}) over the ground state $0^+_1$ and assuming that the ground state is mainly related by $E0$ transition to the $0^+_2$ state we obtain
\begin{eqnarray}
\label{eq:twelve}
| \langle 0^+_2 |\beta^2 | 0^+_1 \rangle |^2 \le \frac{2\hbar^2}{B_0} \langle 0^+_1 |\beta^2 | 0^+_1 \rangle \frac{1}{E(0^+_2)},
\end{eqnarray}
where $E(0^+_2)$ is the excitation energy of the second $0^+$ state. The sign of inequality in~(\ref{eq:twelve}) appears because we neglect a contribution into the value of $\langle 0^+_1 | \beta^2 H \beta^2 | 0^+_1 \rangle$
 of the other $0^+$ states higher in energy than $0^+_2$ .
The quantity $\langle 0^+_1 | \beta^2 | 0^+_1 \rangle$ can be expressed with a good accuracy through the $B(E2; 2^+_1 \rightarrow 0^+_1)$ value using the collective model definition of the $E2$ transition operator:
\begin{eqnarray}
\label{eq:thirteen}
 \langle 0^+_1 |\beta^2 | 0^+_1 \rangle = \displaystyle{\frac{5 B(E2; 2^+_1 \rightarrow 0^+_1)}{ \left(\frac{3}{4\pi} Ze R^2 \right)^2}   }.
\end{eqnarray}
Substituting~(\ref{eq:twelve}) and~(\ref{eq:thirteen}) into~(\ref{rho}) we obtain
\begin{eqnarray}
\label{eq:fourteen}
\rho^2 (0^+_2 \rightarrow 0^+_1) \le \frac{\hbar^2}{B_0}  \frac{1}{E(0^+_2)} \displaystyle{\frac{10 B(E2; 2^+_1 \rightarrow 0^+_1)}{ e^2 R^4} }.
\end{eqnarray}
In our calculations the value of $\dfrac{\hbar^2}{B_0}$ was fixed as  5 keV in order to reproduce the experimental value of $E(0^+_2)$. Substituting this value  and the calculated values of $E(0^+_2)$ and $B(E2; 2^+_1 \rightarrow 0^+_1)$ into (\ref{eq:fourteen}) we obtain that
\begin{eqnarray}
\label{eq:fifteen}
\rho^2 (0^+_2 \rightarrow 0^+_1) \le 0.005
\end{eqnarray}
in correspondence with the result given in Table \ref{table:one}.

This result means that we can not  exclude that the pairing vibrational or some other modes   play an important role in the description of the E0 transitions.

\begin{figure*}[htpb]
a)
\includegraphics[width=0.35\textwidth]{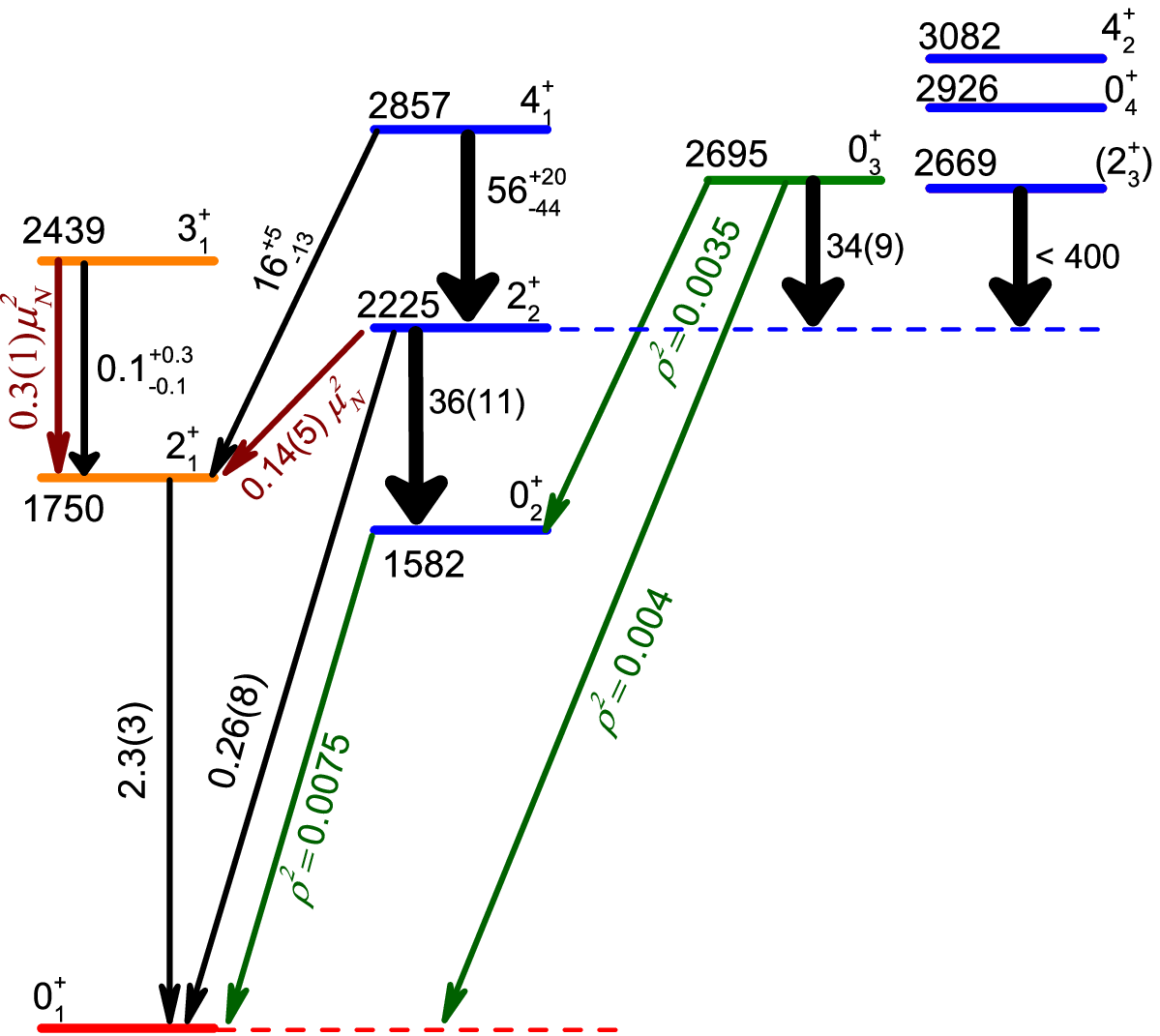}
\hspace*{1cm}
b)
\includegraphics[width=0.35\textwidth]{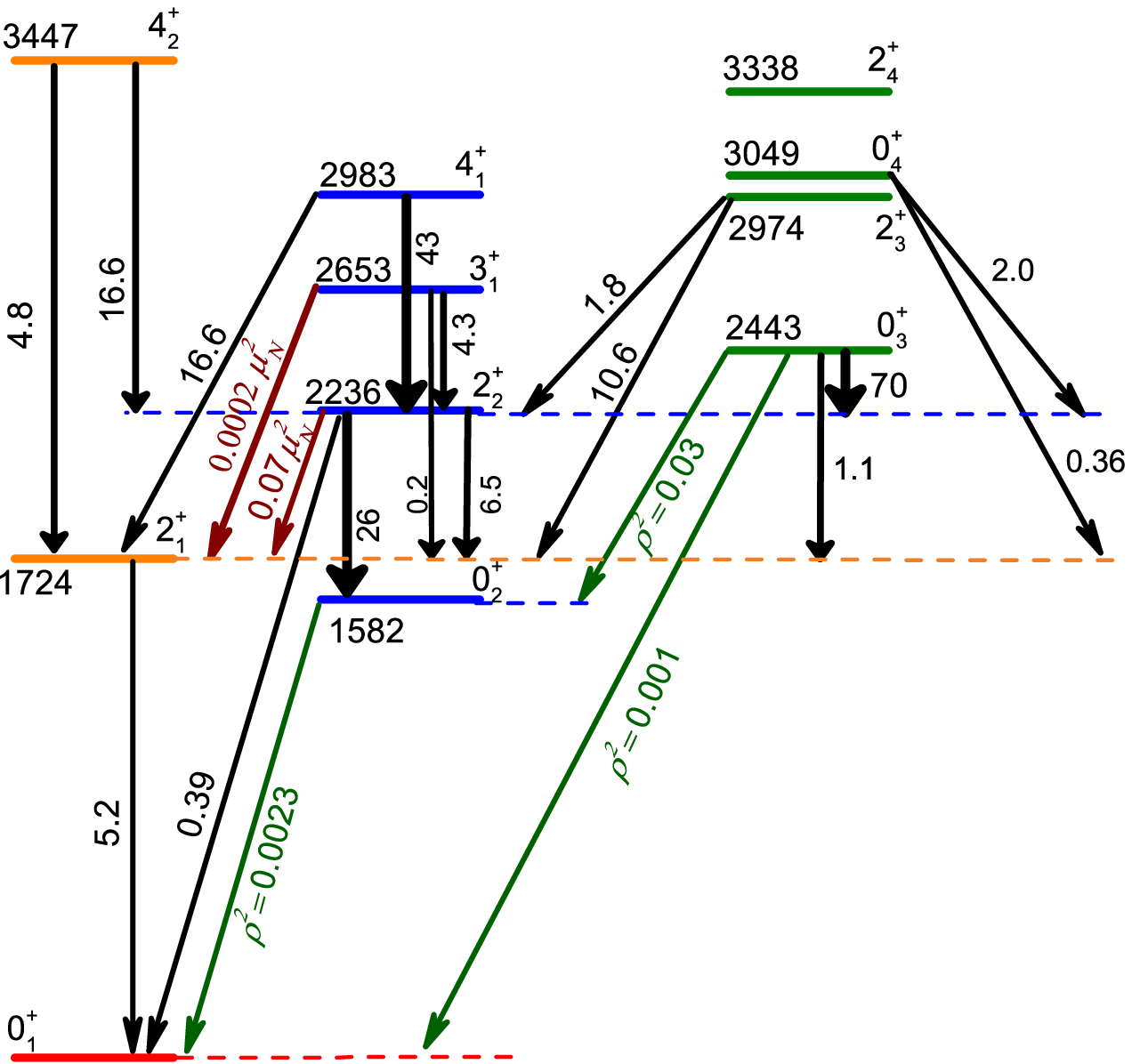}
\caption{\label{f2exp} Experimental (a),~\cite{nndc1,Witt}) and calculated (b) low-energy level scheme of positive-parity states of $^{96}$Zr.
Excitation energies are given in keV, B(E2) transitions are given in W.u.}
\end{figure*}
All experimental data on low-lying excited states of $^{96}$Zr are presented in Fig.~\ref{f2exp}a and the corresponding calculation results are shown in Fig.~\ref{f2exp}b. 
In both figures, states having very large spherical or deformed components are highlighted in two separate columns on the left. In Fig. \ref{f2exp}b the division is based on the results of wave functions weights $W_{In}$ calculation which are shown in Table III. In contrast to the results~\cite{Witt} presented in Fig.~\ref{f2exp}a we placed the $4^+_2$ state among the spherical and $3^+_1$ state among the deformed ones  based on the results shown in Table \ref{table:three}.
\begin{table}[tbh]
\centering
\setlength\aboverulesep{0pt}\setlength\belowrulesep{0pt}
\setcellgapes{3pt}\makegapedcells
\caption{The calculated weights $W_{In}$ in the spherical minimum of the considered states of $^{96}$Zr.}
\label{table:three}
\begin{tabular}{l|c||l|c||l|c}
\hline
State  & $W_{In}$ & State & $W_{In}$ & State & $W_{In}$  \\
\hline
$ 0^+_1$ & 0.985 & $0^+_2$ & 0.136 & $0^+_3$ & 0.292    \\
$ 2^+_1$ & 0.772 & $2^+_2$ & 0.182 & $2^+_3$ & 0.289    \\
$ 4^+_2$ & 0.636 & $4^+_1$ & 0.139 & $0^+_4$ & 0.202    \\
         &       & $3^+_1$ & 0.042 & $2^+_4$ & 0.464   \\

\hline
\end{tabular}
\end{table}

\section{Conclusion}

We have studied a possibility to describe the properties of the low-lying collective quadrupole states of $^{96}$Zr basing on the Bohr collective Hamiltonian. Both $\beta$ and $\gamma$ shape collective variables are included into consideration. The  $\beta$-dependence of the potential energy is fixed to describe the experimental data in a best possible way. However, a $\gamma$-dependence of the potential is introduced in a simple way favoring axial symmetry at large $\beta$. The resulting potential has two minima, spherical and deformed, separated by a barrier. The inertia tensor is taken in a diagonal form with the same values for both $\beta$- and $\gamma$-vibrational modes. However, the rotational inertia coefficient is taken to be 4 times smaller than the vibrational one in order to reproduce the excitation energy of the $2^+_2$ state. Rather good agreement with the experimental data is obtained for the excitation energies and the E2 transition probabilities. The calculated B(M1; $2^+_2\rightarrow 2^+_1$) value is two times smaller than the experimental value. Consideration of the $3^+_1\rightarrow 2^+_1$
M1 transition probabilities indicates the importance of knowledge of the microscopic structure of that part of the collective state wave function that is localized in the spherical minimum. At the same time our calculations show that the wave function  of the $3^+_1$ state is localized mainly in the deformed minimum. Thus, our calculations indicate the problem in the description of the properties of the $3^+_1$ state of $^{96}$Zr in the framework of the Geometrical Collective Model.
The calculated value of $\rho^2(0^+_2\rightarrow 0^+_1)$ is around three times smaller than the measured value. This indicates an influence of the other degrees of freedom that is not included in the present consideration.


\section{Acknowledgments}
The authors express their gratitude to the RFBR (grant № 20--02-00176) and Heisenberg--Landau Program for support.
One of the authors, N.P., thanks A. Leviatan, T. Otsuka, V. Werner, and T. Beck for discussion and gratefully acknowledges support by the DFG under grant N SFB1245 and by the BMBF under grant Nos. 05P19RDFN1 and 05P18RDEN9.

\end{document}